\documentclass[11pt,a4paper]{article}
\usepackage{jheppub}
\usepackage{amsmath}
\usepackage[most]{tcolorbox}
\usepackage{dsfont}
\usepackage{ulem}
\usepackage{natbib}
\usepackage{xcolor}
\usepackage[hang,flushmargin]{footmisc}
\usepackage{tikz-cd}
\usepackage{enumitem}
\usepackage{amsfonts,amssymb, amscd,amsmath,latexsym,amsbsy, bm}
\usepackage{stmaryrd}
\usepackage{todonotes}
\usepackage{float}

\makeatletter\renewcommand{\@biblabel}[1]{#1.}\makeatother

\newtcolorbox{empheqboxed}{colback=gray!20,
 colframe=white,
 width=\textwidth,
 sharpish corners,
 top=0mm,
 bottom=0pt
}

\title{Arithmetic selection rules in dispersionless Hamiltonian systems}

\author{Fatma Aydogmus$^a$ and Mustafa Mullahasanoglu$^{b,c}$ }
\affiliation{
$^a$ Department of Physics, Istanbul University, 34134 Istanbul, Türkiye\\[-0.4cm]

$^b$ Department of Physics, Bogazici University, 34342, Istanbul, Türkiye\\[-0.4cm]

$^c$ Feza Gursey Center for Physics and Mathematics, Bogazici University, Istanbul, Türkiye\\[-0.4cm]
}
\emailAdd{fatmaa@istanbul.edu.tr}
\emailAdd{mustafa.mullahasanoglu@std.bogazici.edu.tr}

\abstract{In this work, we derive selection rules imposed by Liouville integrability conditions for monomial charge densities with arbitrary powers. For a certain monomial Hamiltonian system, the selection rules reduce to a negative Pell equation, and its solutions generate an infinite set of integrals of motion that are mutually in involution.

Furthermore, we study the correspondence between combinatorial polynomial sequences and Liouville integrable Hamiltonian field theories in $1+1$ dimensions. 
We show that the Motzkin system coincides with the dispersionless
limit of the Levi system, while the binomial system is equivalent to the dispersionless derivative nonlinear Schrödinger equation.
Additionally, we show that the binomial Hamiltonian model admits a reduction to the inviscid Burgers equation and its higher-order charges generate generalized Burgers-type conservation laws. 
}

\keywords{Hamiltonian system, Liouville integrability, Levi system, derivative nonlinear Schrödinger equation, Burgers equation, Pell equation, dispersionless systems.}

\begin{document}
\maketitle

\section{Introduction}

Integrable field theories occupy a special position between linear dynamics and generic nonlinear evolution. Although their equations can be strongly nonlinear, they possess sufficiently many conserved quantities to constrain the motion and, in favourable cases, to permit an exact solution. In Hamiltonian language and in the Liouville integrability sense, the relevant condition is the existence of an infinite set of functionally independent charges in involution. Systems of hydrodynamic type provide a particularly useful setting for studying this mechanism because their Hamilton equations are first-order quasilinear partial differential equations and may often be expressed as conservation laws \cite{Arnold, Faddeev:1987ph, Retore:2021wwh, Driezen:2021cpd, olver, dubrovin}.

A different route to such Liouville integrable systems begins with combinatorics. In \cite{Gahramanov:2017att}, Motzkin polynomials were used as densities of conserved charges in a Hamiltonian field theory. The coefficients in polynomials count lattice paths, while a field-theoretic parametrization converts the polynomial sequence into an infinite family of integrals of motion. This observation is supported by the binomial Hamiltonian case in \cite{Mullahasanoglu:2023djn}. Therefore, we investigate polynomial sequences that generate integrable Hamiltonian systems, and what their polynomial structure can tell us about their integrable structure.

The nonlinear Schrödinger equation (NLS) and derivative nonlinear Schrödinger equation (DNLS) are standard examples of dispersive
integrable systems \cite{KaupNewell, shabat} in nonlinear evolution. The Levi system \cite{Levi1981, MikShaYam87} belongs to the same network, and it is
related to NLS by a B\"acklund transformation and to DNLS by a
differential substitution \cite{KunduStramppOevel1995, Fan2001}. 
The purpose of the present work is to find the connections between known dispersive models and the combinatorial models. 
We explicitly show that the Motzkin model corresponds to the dispersionless limit of the Levi system, whereas the binomial model reduces to the dispersionless DNLS, equivalently, the Kaup–Newell system.

The second observation in this work is that the coupled fields possess a closed scalar conservation law. The composite field form of the equations of motion of the binomial model satisfies a Burgers equation \cite{Bonkile:2018bjm, Lou:2024fhu, Reshetikhin:2015eja}. Thus, the same variable that generates every conserved charge also diagonalizes an essential part of the nonlinear evolution. Adding diffusion gives a natural regularization which is mapped to the heat equation by the Cole-Hopf transformation.

A central result of this study is the derivation of arithmetic selection rules from the Liouville integrability conditions. For a monomial Hamiltonian with fixed exponents, the pairwise exponents governing all integrals of motion are shown to satisfy a negative Pell-type Diophantine equation. The infinite family of solutions to this Pell equation guarantees the existence of infinitely many mutually commuting constants of motion. Consequently, the resulting equations of motion define a novel dispersionless class of integrable systems.

The paper is organized as follows. Section~\ref{sec:framework} introduces the Hamiltonian conventions and briefly reviews the Liouville integrability. Section~\ref{sec:motzkin} revisits the Motzkin model and shows its trivial extension to match with the dispersionless Levi system. Then Section~\ref{sec:binomial} develops the binomial model and its connection with dispersionless DNLS systems and the inviscid Burgers equation. Section~\ref {sec:selection} introduces selection rules coming from two conditions of Liouville integrability for monomial charges.
Finally, Section~\ref {sec:conclusion} summarizes the results in the correspondence between combinatorics and Liouville integrability of a $1+1$ dimensional field theory and discusses open problems.

\section{Liouville integrability}
\label{sec:framework}
We briefly revisit Liouville integrability and the Hamiltonian framework in $1+1$ dimensional classical field theory. Basic ingredients of a theory are a real scalar field $\phi(x,t)$ and its canonical momentum $\pi(x,t)$, with equal-time Poisson bracket
\begin{equation}
 \{\phi(x),\pi(y)\}=\delta(x-y), \qquad
 \{\phi(x),\phi(y)\}=\{\pi(x),\pi(y)\}=0 \:.
 \label{canonicalPB}
\end{equation}
Equivalently, one can also define the Poisson bracket for two functionals $F[\phi,\pi]$ and $G[\phi,\pi]$ as 
\begin{equation}
 \{F,G\}=\int dx\left(
 \frac{\delta F}{\delta\phi}\frac{\delta G}{\delta\pi}
 -\frac{\delta F}{\delta\pi}\frac{\delta G}{\delta\phi}
 \right),
 \label{functionalPB}
\end{equation}
where the functional derivative of $F[q(x)]$ with respect to $q$ is the following
\begin{equation}
 \frac{\delta F[q(x)]}{\delta q}
 =\frac{\partial f}{\partial q}
 -\frac{d}{d x}\left(\frac{\partial f}{\partial q'}\right).
 \label{EulerDerivative}
\end{equation}
with the notation prime meaning total derivative with respect to $x$. We assume that the integrations by parts bring a boundary term that vanishes due to the region of integration, like periodic fields or sufficiently rapid decay at spatial infinity. This contribution is essential with open boundaries, but the assumptions remove boundary fluxes \cite{Gieres:2021ekc}.

The Hamiltonian of a model $H[\phi,\pi]$ yields the equations of motion
\begin{equation}
 \frac{\delta H}{\delta \pi}
 =\dot{\phi}\:, \qquad \frac{\delta H}{\delta \phi}
 =-\dot{\pi}\:.
 \label{Heom}
\end{equation}
A Liouville integrable model requires two conditions simultaneously: a number of integrals of motion for its degrees of freedom and the involution of those integrals. When a sufficient set of conserved charges exists, the integrability property is ultimately determined by whether these integrals are in involution, meaning that the Poisson bracket of any two integrals of motion vanishes.

A model in a classical field theory consists of infinitely many degrees of freedom. One of the crucial tasks is the identification of corresponding constants of motion in a theory. Therefore an infinite set of conserved functionals $I_n\equiv I_n[\phi,\pi]$ is required
\begin{equation}
 \frac{d I_n}{dt}=0\:.
 \label{LiouvilleConditions1}
\end{equation}
If one can also show that these conserved quantities satisfy the involution condition
\begin{equation}
\{I_n,I_m\}=0\:,
 \label{LiouvilleConditions2}
\end{equation}
The two principal conditions of Liouville integrability are satisfied. 

We note that the integrals of motion are expected to be functionally independent. Otherwise, integrals of motion become entirely redundant and do not characterize distinct symmetries. This means that an integral of motion cannot be written in terms of any other, and the variational derivatives of integrals of motion are linearly independent. 

\section{The Motzkin model}
\label{sec:motzkin}
The Hamiltonian of the Motzkin model is
\begin{equation}
 H=\int dx\left(\pi^2\phi'+\pi\phi'^2\right)\:.
 \label{MotzkinHamiltonian}
\end{equation}
This Hamiltonian, under the limit $\pi>>\phi$, gives a Hamiltonian \cite{Akhmedov:2010mz} (see also, e.g., \cite{Akhmedov:2010sw}) for the simplest case, matrix scalar field theory, which is written to describe RG flow equations. In a field theory under a change of scale \cite{Polchinski:1983gv, Bervillier:2004mf}, the exact renormalization group studied by the Polchinski equation allows the investigation of dynamics of operators. The equations coming from this reduced Hamiltonian \cite{Akhmedov:2010mz} describe shock waves, and the governing equations are known to be integrable. 

Hamilton's equations yield
\begin{equation}
\dot\pi=2\pi\pi'+2\pi'\phi'+2\pi\phi'',
 \qquad
\dot\phi'=2\phi'\phi''+2\pi'\phi'+2\pi\phi''.
 \label{MotzkinEOM}
\end{equation}
We represent the equations of motion of the Motzkin model in the hydrodynamic system notation by introducing $u=\pi$ and $v=\phi'$ variables, and it yields (see a hydrodynamic perspective for the Born-Infeld equations in \cite{arik:1989, dubrovin})
\begin{equation}
 u_t=(u^2+2uv)_x, \qquad v_t=(v^2+2uv)_x.
 \label{MotzkinHydro}
\end{equation}
Hamiltonian formulations of perfect-fluid systems provide a broader continuum-mechanical context for such hydrodynamic variables and conservation laws \cite{Jackiw:2004nm}.
The model is called the Motzkin model since the infinite set of integrals of motion construction is based on the Motzkin polynomials \cite{Gahramanov:2017att}
\begin{equation}
 I_n=\int dx P_n.
 \label{MotzkinI}
\end{equation}
The coefficients of the polynomial sequence encode Motzkin paths and are related to the Motzkin sequence A001006 \cite{OEIS,motzin}
\begin{equation}
 P_n(\alpha,\beta)=\sum_{k=1}^{\lfloor n/2\rfloor}
 \alpha^k\beta^{n-2k}t_{n,k},
 \qquad
 t_{n,k}=\frac{(n-2)!}{(n-2k)!k!(k-1)!}\:,
 \label{MotzkinPolynomial}
\end{equation}
where the field and momentum substitution is imposed to turn these polynomials into charge densities
\begin{equation}
 \alpha=\pi\phi', \qquad \beta=\pi+\phi'\:.
\end{equation}
With $P_1=\beta$, some polynomials and Motzkin number coefficients are
\begin{align}
 P_2&=\alpha, & P_3&=\alpha\beta, &
 P_4&=\alpha\beta^2+\alpha^2,\nonumber\\
 P_5&=\alpha\beta^3+3\alpha^2\beta, &
 P_6&=\alpha\beta^4+6\alpha^2\beta^2+2\alpha^3\:.
\end{align}
Therefore, the Hamiltonian of the Motzkin model corresponds to the third polynomial
\begin{equation}
 H=\int dx P_3.
\end{equation}
One can verify that all integrals of motion defined by \eqref{MotzkinI} are conserved under time derivative. Also, the Poisson brackets of arbitrarily chosen two polynomial charges are zero. That is, the model constructed by the combinatorial sequence Motzkin is integrable in the Liouville sense.

This example of producing an integrable model in classical field theory demonstrates how a combinatorial sequence can organize an infinite hierarchy of field-theory charges. It also motivates the simpler correspondence between combinatorics and Liouville integrability.
\subsection{The dispersionless Levi system}
Here we express the Motzkin model as a dispersionless form of the Levi system. The Levi system is
\begin{equation}\label{Levi}
    \begin{aligned}
 u_t&=u_{xx}+
 \left(u^2+2uv+k u\right)_x,
\\
 v_t&=-v_{xx}+
 \left(v^2+2uv+k v\right)_x.
\end{aligned} 
\end{equation}
The opposite signs of \(u_{xx}\) and \(v_{xx}\) are characteristic of
two-field NLS-type systems. 

First, we show that shifting the $\beta$ variable by a constant $k$ does not affect the structure and preserves the Liouville integrability of the Motzkin model. In the field theory framework, the shift parameter \(k\) in $\beta$ enters as a linear contribution to the two fluxes. From the definition of the Motzkin polynomial \eqref{MotzkinPolynomial}, we can easily see that the polynomials take the form of linear combinations. 

Here we present more explicitly that this shift is not more than binomial combinations of usual integrals of motion. Let's denote $\Bar{\beta}=\beta+k$ and consider several polynomials
\begin{align}
 \Bar{P}_2&=\alpha, & \Bar{P}_3&=\alpha\Bar{\beta}, &
 \Bar{P}_4&=\alpha\Bar{\beta}^2+\alpha^2,\nonumber\\
 \Bar{P}_5&=\alpha\Bar{\beta}^3+3\alpha^2\Bar{\beta}, &
 \Bar{P}_6&=\alpha\Bar{\beta}^4+6\alpha^2\Bar{\beta}^2+2\alpha^3\:.
\end{align}
One can observe that this redefinition is not a nontrivial set of sequences but a linear composition of polynomials in the binomial sense 
\begin{align}
 \Bar{P}_2&=P_2, & \Bar{P}_3&=P_3+kP_2,\qquad
 \Bar{P}_4=P_4+2kP_3+k^2P_2,\nonumber\\
 \Bar{P}_5&=P_5+3kP_4+3k^2P_3+k^3P_2, &
 \Bar{P}_6&=P_6+4kP_5+6k^2P_4+4k^3P_3+k^4P_2\:.
\end{align}
Therefore, the shift does not produce a new integrable model since the functional independence is the same as the $k=0$ case.

To exploit the Levi system, consider the Hamiltonian of the extended Motzkin model 
\begin{equation}
 H=\int dx\left(\pi^2\phi'+\pi\phi'^2+k\pi\phi'\right)\equiv \int dx \Bar{P}_3.
 \label{eMotzkinHamiltonian}
\end{equation}
The equations of motion of the Motzkin model take the following form 
\begin{equation}
 u_t=(u^2+2uv+ku)_x, \qquad v_t=(v^2+2uv+kv)_x.
 \label{eMotzkinHydro}
\end{equation}
This expression corresponds precisely to the dispersionless formulation of the Levi system.
 
We demonstrate that the zero-curvature representation of the Motzkin model can be identified with the Abelian form of the Levi system.  
To make this correspondence explicit, we introduce two matrices
\begin{equation}
  U=
 \begin{pmatrix}
 s-\lambda&-v\\
 u&\lambda-s
 \end{pmatrix},\qquad 
  V
 =2(\lambda+s) U
 +2
 \begin{pmatrix}
 \frac12(u-v)&v\\
 u&\frac12(v-u)
 \end{pmatrix}_x.
\label{eq:levi-UV}
\end{equation}
where $\lambda$ is a spectral parameter and \(2s=u+v+k/2\). 

The off-diagonal entries of the zero curvature condition reproduce the Levi system 
\[
 U_t= V_x+[ V, U].
\]
If we reduce the non-Abelian part of the zero curvature condition by removing the derivative-dependent term, the second term of matrix $V$, we end up with the equations of motion of the extended Motzkin model. The naive matrix zero-curvature limit becomes Abelian and does not retain the essential spectral information of the parent Levi system.

In other words, we introduce a parameter \(\epsilon\) in the second-order terms of the Levi equations (\ref{Levi}) and take the dispersionless limit as \(\epsilon\to0\) to deduce (\ref{eMotzkinHydro}). Together with the limit, the choice $k=0$  gives exactly (\ref{MotzkinHydro}) and it is already the same even for $k\neq 0$.

Here we note that the Levi system is related to the NLS system by the B\"acklund transformation
\begin{equation}
 pq=uv-v_x,\qquad
 \frac{p_x}{p}=u+v+k.
\label{eq:levi-to-nls}
\end{equation}
The transformation resembles the field-polynomial reparameterization, but the second equation implies that this transformation is not a point transformation.  Once \(p\) is reconstructed, the first
equation determines \(q\).
The compatibility of these relations with the Levi time evolution
produces the NLS system, up to the conventional normalizations of
\(x,t,p,q\).

By the differential substitution, the Levi system can be written in DNLS form
\begin{equation}
 u=\frac{b_x}{b}-\frac{ab}{2},
 \qquad
 v=-\frac{ab}{2}.
\label{eq:levi-to-dnls}
\end{equation}
Consequently, suppressing derivatives before applying the
transformation does not necessarily commute with applying the
transformation first.  Dispersionless reductions must therefore be
handled at the level of the equations rather than inferred only from
the full dispersive transformation.

Even when the limiting zero-curvature equation becomes Abelian, this
does not contradict the integrability of the parent Levi system.  It
only means that the selected limit has discarded part of the spectral
geometry.  A proper dispersionless Lax representation often uses a
Poisson bracket in a momentum variable instead of an ordinary matrix
commutator \cite{Kamchatnov, KAMCHATNOV2024}.

\section{The binomial model}
\label{sec:binomial}
In this part, we revisit the Liouville integrable binomial model \cite{Mullahasanoglu:2023djn}. The combinatorial structure of the Motzkin model motivated searching for the density of charges $P_n=(\alpha+\beta)^n$, and the corresponding model is constructed in \cite{Mullahasanoglu:2023djn}. 
The integrals of motion and the Hamiltonian of the binomial model are introduced as
\begin{equation}
 I_n=\int dx\,(c\pi\phi^\prime+a\pi+b\phi^\prime)^n, \qquad H\equiv I_2=\int dx\,(c\pi\phi^\prime+a\pi+b\phi^\prime)^2,
 \label{BinomialHierarchy}
\end{equation}
where $a,b,c\in \mathbb{R}$ are constants. The corresponding Lagrangian density of the model is
\begin{align}
    \mathcal{L}= \frac{\dot{\phi}^2}{4(c\phi^\prime+a)^2}-\frac{b\phi^\prime\dot{\phi}}{(c\phi^\prime+a)} \:.
\end{align}
Lagrangian formulations have also been developed for other dispersionless integrable hierarchies, including the dispersionless KdV hierarchy \cite{Choudhuri_2007}.
As seen in the Lagrangian and the Hamiltonian, the parameters $c$ and $a,b$ distinguish into two sectors of the binomial model. Thus, keeping the parameters is dynamically important, and setting them to unity too early hides inequivalent limits such as $c=0$, $a=0$, or equivalently $b=0$. This will be more explicit in the composite form of the equation of motion below. 

In terms of $u=\pi$ and $v=\phi'$, the equations of motion become
\begin{equation}
 u_t=2\big[(cuv+au+bv)(cu+b)\big]_x,
 \qquad
 v_t=2\big[(cuv+au+bv)(cv+a)\big]_x.
 \label{binomequations}
\end{equation}
One can also immediately observe that
\begin{align}
 \frac{dI_n}{dt}
 =n\int dx\,P^{n-1}P_t=\int dx\,\left[
 \frac{6nc}{n+1}P^{n+1}+4abP^n
 \right]_x=0,
\end{align}
where $P=c\pi\phi^\prime+a\pi+b\phi^\prime$.
The total derivative with respect to $x$ will vanish due to the boundary conditions, and it displays the local conservation law and its associated flux.

The involution condition of Liouville integrability is substituted into \eqref{functionalPB} by elementary rearrangement
\begin{align}
 \{I_n,I_m\}
 =\int dx\,\Bigg[
 \frac{(m-n)mn}{m+n-1}cP^{m+n-1}
 +\frac{(m-n)mn}{m+n-2}abP^{m+n-2}
 \Bigg]_x=0.
 \label{InvolutionResult}
\end{align}
The exceptional $m=n=1$ case already vanishes directly. Thus two charges from the set of infinitely many integrals of motion are pairwise in involution. Together with the infinite conserved quantities in \eqref{BinomialHierarchy}, this establishes the formal Liouville integrability of the binomial model.

As a shifted form of the Motzkin model, the integrability of the binomial model is preserved. At this level, the shift becomes trivial, but it will reveal novel integrable models due to the factorization of the densities. So one can rewrite the charge densities as a combination of polynomials by adding a constant term
\begin{equation}
 \Bar{P}=P+k,
 \qquad I_n=\int dx\,\Bar{P}^n.
 \label{ShiftedHierarchy}
\end{equation}
With the $H=I_2$ Hamiltonian, the Lagrangian density becomes
\begin{align}
   H=\int dx\,(c\pi\phi'+a\pi+b\phi'+k)^2,
 \qquad \mathcal{L}= \frac{\dot{\phi}^2}{4(c\phi^\prime+a)^2} -\frac{\dot{\phi}(b\phi^\prime+k)}{(c\phi^\prime+a)} \:.
\end{align}
The equations of motion are the same except replacing $P$ with $\Bar{P}$.
As above, the charges are also conserved,
\begin{align}
 \dot I_n
 &=\int dx\,\left[
 \frac{6nc}{n+1}\Bar{P}^{n+1}+4(ab-ck) \Bar{P}^n
 \right]_x=0\:,
 \label{ShiftedConservation}
\end{align}
and the Poisson bracket calculation vanishes
\begin{align}
 \{I_n,I_m\}=\int dx\,\Bigg[
 \frac{(m-n)mn}{m+n-1}c\Bar{P}^{m+n-1}+\frac{(m-n)mn}{m+n-2}(ab-ck)\Bar{P}^{m+n-2}
 \Bigg]_x=0.
 \label{ShiftedInvolution}
\end{align}
Thus, a constant shift term $k$ does not destroy the involution. It changes the coefficient of the lower-power flux from $ab$ to $ab-ck$. 

We want to factorize the iterated binomial density so that it admits two affine factors
\begin{equation}
 K=(c_1\pi+c_2)(c_3\phi'+c_4),
 \label{Factorization}
\end{equation}
where $c_i\in \mathbb R$ are related to $a,b,c,k$. However, in this factorization, we recognize $ab-ck=0$ in terms of the $c_i$ variables. That means we miss the contribution of a lower-power term to the conservation law. Therefore, we should consider $\Bar{K}=K+c_5$ to be consistent with the number of free parameters in the binomial framework.  

\subsection{The dispersionless DNLS}
The derivative nonlinear Schr\"odinger (DNLS) system, in which the nonlinearity contains a derivative, which is why it is called a derivative NLS equation, also called the Kaup-Newell system \cite{KaupNewell}, is
\begin{equation}
 u_t=u_{xx}+2(u^2v)_x,\qquad
 v_t=-v_{xx}+2(uv^2)_x.
\label{eq:dnls1}
\end{equation}
The binomial model has the equation \eqref{binomequations} and we reduce by
affine field shifts (or $c=1,a=b=0$), a moving frame, and a rescaling to
\[
 u_t=2(u^2v)_x,\qquad v_t=2(uv^2)_x.
\]
This is the dispersionless nonlinear part of the DNLS system. In other words, we insert a dispersion parameter into DNLS and, at \(\epsilon=0\), and after an unimportant rescaling of \(t\), this
is precisely the equations of motion of the binomial model.  


There are three DNLS equations, and they are related to each other by nontrivial transformations. Namely, they are connected by nonlocal gauge transformations \cite{KunduStramppOevel1995, Fan2001}. Locality is not preserved term by term in the transformations even though the transformed equations are
local. The Chen-Lee-Liu \cite{Chen_1979,Olver1998} or DNLS-II system with independent-field convention is
\begin{equation}
 u_t=u_{xx}+2uvu_x,\qquad
 v_t=-v_{xx}+2uvv_x.
\label{eq:dnls2}
\end{equation}
The Gerdjikov-Ivanov \cite{GerdjikovIvanov,Kudryashov2023bifurcation} or DNLS-III system has the schematic form
\begin{align}
 u_t&=u_{xx}-2u^2v_x+\frac12 u^3v^2,\\
 v_t&=-v_{xx}-2v^2u_x-\frac12 u^2v^3.
\label{eq:dnls3}
\end{align}
Note that factors like signs
and constants vary in the literature because of the choice of different real or complex times and different conjugation reductions \cite{shabat}.


Here we remark that the new model \eqref{newequations} is not in the same family, and this can be recognized by the power of the fields. The idea of this combinatorics and Liouville integrability construction may reveal various models, and their dispersion part should be studied in more detail. 
\subsection{The Burgers equation}
\label{sec:Burgers}
The equations of motion \eqref{binomequations} can be expressed in the composite field configuration $P=cuv+au+bv$, then the equation of motion takes the following form
\begin{equation}
 P_t=(3cP^2+4abP)_x\:.
 \label{PcompositeBinom}
\end{equation}
Here we recognize that the composite form of the fields becomes the inviscid Burgers equation for the $a=b=0$ case. Without any fixing, the composite scalar dynamics is still of the inviscid Burgers equation \cite{Bonkile:2018bjm}. The change of field variable $U=-3c\Bar{P}-2(ab-ck)$ with $k$ shift yields
\begin{equation}
U_{t}+(U^2)_{x}=0 .
\end{equation}
The original two-component problem therefore has a scalar projection. Notice that this does not by itself reconstruct $u$ and $v$ separately; it determines the combination $P$ that generates the complete charge hierarchy.

The case $c=0$ must be treated separately because the rescaling is singular, then reduces to the linear advection equation, which does not exhibit nonlinear wave steepening.

A standard regularization of the scalar conservation law is obtained by adding diffusion,
\begin{equation}
 \Bar{P}_t=\partial_x(3c\Bar{P}^2+4(ab-ck)\Bar{P})+\nu \Bar{P}_{xx},
 \qquad \nu>0.
 \label{ViscousP}
\end{equation}
This is an additional dissipative deformation and should not be confused with the original Hamiltonian flow. Under $U=-3c\Bar{P}-2(ab-ck)$, the equation \eqref{ViscousP} becomes Burgers equation
\begin{equation}
 U_t+(U^2)_x=\nu U_{xx}.
 \label{ViscousBurgers}
\end{equation}
We show that the composite field $\bar P$ respects the Cole-Hopf substitution
\begin{equation}
 \bar P
 =\frac{1}{3c}\left(\nu\frac{\phi_x}{\phi}-2(ab-ck)\right).
 \label{KfromHeat}
\end{equation}
maps \eqref{ViscousBurgers} to the linear heat equation
\begin{equation}
 \phi_t=\nu\phi_{xx}.
 \label{HeatEquation}
\end{equation}
The small $\nu$ limit selects the entropy solution of the inviscid equation.

We showed that the Hamiltonian $H=I_2$ of the binomial model produces the quadratic Burgers-type flux discussed above. Now we search for the existence of the entire set of commuting integrals of motion, which suggests a stronger statement. Every higher charge may itself be chosen as a Hamiltonian and generates another compatible conservation law flow. This embeds the quadratic model into a infinite set of charges with polynomial fluxes of arbitrary degree. Related constructions of commuting flows for two-component dispersionless PDEs are studied in \cite{MANGANARO2024}.

We choose
\begin{equation}
 H=I_N=\int dx\,P^N,\qquad N\geq2.
 \label{HigherHamiltonian}
\end{equation}
Hamilton's equations are
\begin{equation}
 u_{t}=N\big[P^{N-1}(cu+b)\big]_x,
 \qquad
 v_{t}=N\big[P^{N-1}(cv+a)\big]_x.
 \label{HigherTwoFieldFlow}
\end{equation}
The flows are mutually compatible at the Hamiltonian level. Indeed,
\begin{equation}
 \frac{\partial I_M}{\partial t}
 =\{I_M,I_N\}=0
 \label{HigherFlowCommutativity}
\end{equation}
for every $M, N$ and subject to the usual regularity and functional-independence qualifications.

We use the composite field form $P$ and a direct calculation yields
\begin{align}
 P_{t}
 =\left[(2N-1)cP^N
 +2N ab P^{N-1}\right]_x.
 \label{GeneralScalarLaw}
\end{align}
In the conventional scalar conservation-law form $w_t+f(w)_x=0$, the field $w$ is the conserved density and $f(w)$ is its flux function. 

We have already demonstrated that this formulation reproduces the quadratic flux of the inviscid Burgers equation in the case $N=2$. For general $N$, the role of the parameters becomes apparent once more, as the hierarchy \eqref{GeneralScalarLaw} decouples into the purely power conservation laws for $\alpha=c\,uv$ and $\beta=a\,u+b\,v$ corresponding to $a,b=0$ and $c=0$ in $P$, respectively, considered separately
\begin{align}
 \alpha_{t}-(2N-1)c(\alpha^N)_x=0,
 \\
  \beta_{t}-2Nab(\beta^{N-1})_x=0,
\label{PurePowerLaw}
\end{align}
which are generalized inviscid Burgers equations, up to a rescaling and reversal of time. Related Burgers-hierarchy equations and their solitary waves are discussed in \cite{Kudryashov2025Burgers}.

Consequently, the binomial construction yields not only a scalar reduction but also an entire hierarchy of two-variable, first-order dispersionless hydrodynamic systems.

\section{Arithmetic selection rules}
\label{sec:selection}
We investigate novel integrable models by generalizing charge densities as monomials of the field and its conjugate momentum with selection rules. 
We want to observe that arbitrary powers of the field and momentum result in a selection rule replacing the integrability conditions. A model will become a specific choice of powers in the Hamiltonian. This will control integrability by fixing certain powers of integrals of motion.

Let's enlarge the set of candidate charge densities by allowing independent powers of the two factors
\begin{equation}
 I_{i,j}=\int dx\,\pi^i\phi'^j.
 \label{EnrichedCharges}
\end{equation}
Let the Hamiltonian exponents be denoted by two specific choices $i=r$ and $j=s$, where $r,s$ are fixed from now on
\begin{equation}
 H=I_{r,s}=\int dx\,\pi^r\phi'^s.
 \label{EnrichedHamiltonian}
\end{equation}
The canonical Hamilton equations are
\begin{equation}
 \pi_t=s\big[\pi^r\phi'^{s-1}\big]_x,
 \qquad
 (\phi')_t=r\big[\pi^{r-1}\phi'^s\big]_x.
 \label{EnrichedEOM}
\end{equation}
We check that the arbitrary candidate charge along the $H$ flow gives
\begin{align}
 \dot I_{i,j}=\int dx\,
 \pi^{i+r-2}\phi'^{j+s-2} 
 \Big[
 r(is+jr-j)\pi'\phi'+s(is+jr-i)\pi\phi''
 \Big].
 \label{GeneralChargeDerivative}
\end{align}
We aim to make this expression a total derivative of $\pi^{i+r-1}\phi'^{j+s-1}$ precisely, so the condition becomes
\begin{equation}
 \frac{r(is+jr-j)}{i+r-1}
 =\frac{s(is+jr-i)}{j+s-1}.
 \label{ExponentCondition1}
\end{equation}
This is the basic selection rule for conserved monomials $\dot I_{i,j}=0$ when $H=I_{r,s}$. It turns the search for hierarchies into an arithmetic problem for the exponent pairs $(i,j)$.

Instead of checking the explicit Poisson bracket for arbitrary powers, we search for another selection rule vanishing the Poisson bracket of two integrals of motion
\begin{align}
 \{I_{n,m},I_{k,l}\}
 =\int dx\left[\pi^{n+k-1}\phi'^{m+l-1}\left(\frac{(m-l)nk}{n+k-1}+\frac{m(m-1)k-l(l-1)n}{m+l-1}\right)\right]_x\:.
\end{align}
We use the total derivative and cancel the term by the periodic boundary condition. Then we obtain the second selection rule, which is consistent with the first one
\begin{align}
    \frac{(m-l)nk}{n+k-1}=\frac{m(m-1)k-l(l-1)n}{m+l-1} \:.
\end{align}

Therefore, we express Liouville integrability of monomial type models
\begin{equation}
 I_{i,j}=\int dx\,\pi^i\phi'^j\:, \qquad
 H=I_{r,s}=\int dx\,\pi^r\phi'^s.
\end{equation}
in terms of two selection rules:
\begin{itemize}
    \item For certain Hamiltonian $H=I_{r,s}$, the selection rule from $\dot I_{i,j}=0$ becomes
    \begin{equation}
 \frac{r(is+jr-j)}{i+r-1}
 =\frac{s(is+jr-i)}{j+s-1}.
 \label{ExponentCondition}
\end{equation}
    \item In general the involution $\{I_{n,m},I_{k,l}\}
 =0$ yields the second selection rule
 \begin{align}
    \frac{(m-l)nk}{n+k-1}=\frac{m(m-1)k-l(l-1)n}{m+l-1} \:.
    \label{2ndASR}
\end{align}
\end{itemize}
Let's check the consistency of the selection rules with known models.

If one sets Hamiltonian for $(r,s)=(2,1)$, the selection rule \eqref{ExponentCondition1} reads
\begin{equation}
 \frac{2(i+j)}{i+1}=\frac{2j}{j}.
\end{equation}
For positive exponents, it requires $j=1$, while $i$ remains arbitrary. This produces the following integrable model
\begin{equation}
 I_i=\int dx\,\pi^i\phi',
 \qquad H=\int dx\,\pi^2\phi'.
 \label{A2BHierarchy}
\end{equation}
We check the involution in which the arbitrary integrals of motion commute in the Poisson bracket
\begin{align}
 \{I_n,I_m\}
 &=nm\int dx\,
 \pi^{n-1}\phi'\pi^{m-1}\pi'-nm\int dx\,
\pi^{m-1}\phi'\pi^{n-1}\pi'\nonumber\\
 &=0.
\end{align}
Let's recover the binomial model and set variables $(r,s)=(2,2)$, the selection rule becomes
\begin{equation}
 j^2-j=i^2-i, \qquad \text{or} \qquad (j-i)(j+i-1)=0\:.
\end{equation}
For this Hamiltonian, the nontrivial construction of an integrable model needs the diagonal family $i=j$ of constants of motion
\begin{equation}
 I_i=\int dx\,(\pi\phi')^i,
 \qquad H=\int dx\,(\pi\phi')^2.
\end{equation}
We find reduced binomial model for $(r,s)=(2,1)$ and re-derive the binomial model for $(r,s)=(2,2)$. 
\subsection{The Pell equation construction}
\label{sec:Pell}
We have controlled the selection rules, and here we search for a new integrable model. For $(r,s)=(2,3)$, the equations of motion become different and not symmetric under the exchange of $\pi$ and $\phi'$
\begin{equation}
 \pi_t=3\big[\pi^2\phi'^{2}\big]_x,
 \qquad
 (\phi')_t=2\big[\pi^{}\phi'^3\big]_x.
 \label{newequations}
\end{equation}
The total time derivative of an arbitrary integral of motion should vanish, and the first selection rule turns into 
\begin{equation}
 j^2-j=3(i^2-i),\qquad \text{equivalently} \qquad (2j-1)^2-3(2i-1)^2=-2\:.
 \label{PellCondition1}
\end{equation}
We can write the selection rule as a negative Pell equation in the following form \cite{Matthews2000}
\begin{equation}
 x^2-3y^2=-2,\qquad \text{where} \qquad x=2j-1\:, \qquad y=2i-1\:.
 \label{PellCondition2}
\end{equation}
Unlike the previous two cases discussed above, the Pell equation does not permit an arbitrary exponent. It defines a sparse Diophantine sequence solving this Pell equation, including 
$$(i,j)=(1,1), (2,3), (6,10), (21,36), (77,133), (286,495), (1066,1846), (3977,6888),\ldots,$$
or
$$(y,x)=(1,1), (3,5), (11,19), (41,71), (153,265), (571,989), (2131,3691), (7933,13775),\ldots.$$
One can solve \eqref{PellCondition2} for all positive solutions relevant to the Hamiltonian exponents, which are
generated by \cite{Matthews2000}
\begin{equation}
 x_n+y_n\sqrt{3}
 =
 (1+\sqrt{3})(2+\sqrt{3})^n,
 \qquad n=0,1,2,\ldots .
\end{equation}
Multiplication by the fundamental unit $2+\sqrt{3}$ gives the
recurrence. Therefore, the sequence has the following recursion relation and generates an infinite sequence of solutions to the selection rule \cite{Lenstra2002}
\begin{align}
    y_{k+1}=2y_k+x_k\:,\qquad x_{k+1}=3y_k+2x_k \:,
\end{align}
or
\begin{align}
    i_{k+1}=2i_k+j_k-1\:,\qquad j_{k+1}=3i_k+2j_k-2 \:.
\end{align}
Now we verify the second selection rule for the proposed model by the Pell equation and check that the constructed integrals of motion are in involution. 

Substituting the first Pell condition \eqref{PellCondition1} from the first selection rule into the second selection rule \eqref{2ndASR} yields
 \begin{align}
    \frac{m-l}{n+k-1}=3\frac{n-k}{m+l-1} \:.
\end{align}
This can be written as a difference of two Pell equations and, consequently, every two members of the Pell family are in involution
\begin{align}
    m(m-1)-l(l-1)=3[n(n-1)-k(k-1)]\:.
\end{align}
Therefore, both selection rules of the Pell system are satisfied as an integrability condition, and the resulting Pell-type commuting monomial charges appear to be a new dispersionless integrable model.

Consequently, this novel framework associated with the Pell equation suggests that an augmented Hamiltonian may admit distinguished infinite subsequences of polynomial conserved charges. Naturally, the classification problem becomes to determine which Hamiltonian exponent pairs $(r,s)$ yield infinite Diophantine families of charges.

We do not pursue the development of additional integrable models here and instead leave this direction for future work. A systematic treatment of additional models would yield a more conceptual and extensive body of results, detracting from the central focus of this study. Accordingly, we restrict our attention to the established models and one new case and their relationship to the dispersionless Levi system and derivative nonlinear Schrödinger equation.

\section{Conclusion}
\label{sec:conclusion}
The correspondence between combinatorial sequences and integrable Hamiltonian systems facilitates the formulation of a novel integrable model in which the charge densities are represented by polynomial functions. We demonstrate that the two recently introduced Hamiltonian systems, namely the Motzkin and binomial models, are in fact dispersionless reductions of well-known integrable systems. More explicitly the Motzkin model coincides with the dispersionless limit of the Levi system, while the binomial model reduces to the dispersionless derivative nonlinear Schrödinger (DNLS) equation, also known as the Kaup–Newell system. 

In this study, we show that the Liouville integrability conditions turn into selection rules for the powers of the fields in the monomials. When we focus on a certain choice of exponents in the Hamiltonian, we observe that the pairwise exponents of all integrals of motion should satisfy the negative Pell equation. The infinite set of solutions to the Pell equation reveals infinitely many constants of motion which are in involution. The resulting equations of motion for the new model constitute a dispersionless class of integrable differential equations. So dispersive models should be regarded as members of a connected integrable network  \cite{shabat}. Systematic deformations of dispersionless Lax systems and reconstructions of dispersive equations from their dispersionless limits provide frameworks for studying possible dispersive completions \cite{Krynski2023, Ferapontov2009}. Therefore, the new model must use a dispersive deformation and survive equivalence tests against the known NLS/DNLS classifications, including general-form NLS equations and their reductions \cite{Kudryashov2025NLS}. 

The arithmetic selection rules derived from the two principal conditions of Liouville integrability open several directions for further investigation in dispersionless Hamiltonian systems.
For instance, the extension of these models to higher spacetime dimensions is one of them. While the present work is confined to $1+1$ dimensions, one may ask whether analogous integrable structures exist for fields depending on $1+2$ (or more) coordinates. So, by arithmetic selection rules, multidimensional dispersionless systems can be studied through hydrodynamic reductions, and higher conservation laws \cite{Ismailov2011, Ferapontov2025, Dunajski:2008at, Calderbank:2016bqu}. 

The combinatorial perspective and selection rules for the Hamiltonian framework of classical field theories become more suitable with recent machine-learning approaches \cite{Krippendorf_2021, LopesCardoso:2024tol, Fukushima:2026klz, Sipka_2023}. In these works, searching for Lax pairs and connections, conserved quantities, and integrable Hamiltonian families are provided, and these are also complementary tools for testing candidate dispersive deformations of the Pell-type new model. If a Lax construction can be found, it would be interesting to determine whether the dispersionless systems discussed here admit gauge-theoretic realizations as two-dimensional integrable field theories provided by the four-dimensional Chern-Simons theory \cite{Costello:2019tri, Ashwinkumar:2023zbu, Yamazaki:2025yan}. 

In the binomial model, we recognized two distinct sectors with a specific choice of parameters $c=0$ and $a,b=0$. This combinatorial framework for constructing autonomous integrable Hamiltonian systems accommodates constant parameters \(a, b, c, k\). A natural direction is the development of time-dependent generalizations of these models by allowing the conserved quantities themselves to evolve through the time-dependent parameters \(a(t), b(t), c(t), k(t)\).

\section*{Acknowledgments}
We thank Ilmar Gahramanov for his comments on the manuscript. We are grateful to Aslan Salari and Güliz Konca Adiller for the discussions. F. A. and M. M. are supported by the Scientific and Technological Research Council of Turkey (TÜBİTAK) under grant number 125F596.


\bibliographystyle{utphys}
\bibliography{bibtex}

@article{Gahramanov:2017att,
    author = "Gahramanov, Ilmar and Musaev, Edvard T.",
    title = "{Integrability properties of Motzkin polynomials}",
    eprint = "1706.00197",
    archivePrefix = "arXiv",
    primaryClass = "hep-th",
    doi = "10.1063/1.5018372",
    journal = "J. Math. Phys.",
    volume = "61",
    number = "3",
    pages = "033509",
    year = "2020"
}

@article{Fukushima:2026klz,
    author = "Fukushima, Osamu and Shigemura, Tomohiro and Suda, Ryosuke and Tanahashi, Norihiro and Yoshida, Kentaroh",
    title = "{Machine-Learning Search for Lax Connections}",
    eprint = "2608.05146",
    archivePrefix = "arXiv",
    primaryClass = "hep-th",
    reportNumber = "RIKEN-iTHEMS-Report-26 KUNS-3116STUPP-26-302",
    year = "2026"
}

@Article{Kudryashov2023bifurcation,
AUTHOR = {Kudryashov, Nikolay A. and Lavrova, Sofia F. and Nifontov, Daniil R.},
TITLE = {{Bifurcations of Phase Portraits, Exact Solutions and Conservation Laws of the Generalized Gerdjikov–Ivanov Model}},
JOURNAL = {Mathematics},
VOLUME = {11},
YEAR = {2023},
NUMBER = {23},
ARTICLE-NUMBER = {4760},
ISSN = {2227-7390},
DOI = {10.3390/math11234760}
}

@article{Kudryashov2025NLS,
title = {{Nonlinear Schrödinger equations of general form and their exact solutions}},
journal = {Applied Mathematics Letters},
volume = {170},
pages = {109622},
year = {2025},
issn = {0893-9659},
doi = {https://doi.org/10.1016/j.aml.2025.109622},
author = {Nikolay A. Kudryashov and Andrei D. Polyanin}
}

@article{Kudryashov2025Burgers,
  author  = {Kudryashov, N. A.},
  title   = {{Solitary Waves of Burgers Hierarchy Equations}},
  journal = {Computational Mathematics and Mathematical Physics},
  volume  = {65},
  number  = {5},
  pages   = {995--1003},
  year    = {2025},
  issn    = {1555-6662},
  doi     = {10.1134/S0965542525700265}
}

@article{LopesCardoso:2024tol,
    author = "Lopes Cardoso, Gabriel and Mayorga Pena, Daman and Nampuri, Suresh",
    title = "{Classical Integrability in the Presence of a Cosmological Constant: Analytic and Machine Learning Results}",
    eprint = "2404.18247",
    archivePrefix = "arXiv",
    primaryClass = "hep-th",
    doi = "10.1002/prop.202400267",
    journal = "Fortsch. Phys.",
    volume = "73",
    number = "4",
    pages = "2400267",
    year = "2025"
}

@article{Retore:2021wwh,
    author = "Ana L. Retore",
    title = "{Introduction to classical and quantum integrability}",
    eprint = "2109.14280",
    archivePrefix = "arXiv",
    primaryClass = "hep-th",
    doi = "10.1088/1751-8121/ac5a8e",
    journal = "J. Phys. A",
    volume = "55",
    number = "17",
    pages = "173001",
    year = "2022"
}

@article{KaupNewell,
    author = {Kaup, David J. and Newell, Alan C.},
    title = {{An exact solution for a derivative nonlinear Schrödinger equation}},
    journal = {Journal of Mathematical Physics},
    volume = {19},
    number = {4},
    pages = {798-801},
    year = {1978},
    doi = {10.1063/1.523737}
}

@article{Levi1981,
doi = {10.1088/0305-4470/14/5/028},
year = {1981},
publisher = {},
volume = {14},
number = {5},
pages = {1083},
author = {D Levi},
title = {{Nonlinear differential difference equations as Backlund transformations}},
journal = {Journal of Physics A: Mathematical and General}
}

@article{Olver1998,
doi = {10.1007/s002200050328},
year = {1998},
publisher = {},
volume = {193},
number = {2},
pages = {245-268},
author = {Olver, Peter J. and Sokolov, Vladimir V.},
title = {Integrable Evolution Equations on Associative Algebras},
journal = {Communications in Mathematical Physics}
}

@article{Chen_1979,
doi = {10.1088/0031-8949/20/3-4/026},
year = {1979},
volume = {20},
number = {3-4},
pages = {490},
author = {H H Chen and Y C Lee and C S Liu},
title = {{Integrability of Nonlinear Hamiltonian Systems by Inverse Scattering Method}},
journal = {Physica Scripta}
}

@article{GerdjikovIvanov,
year = {1983},
volume = {10},
pages = {130--143},
author = {V. S. Gerdjikov and M. I. Ivanov},
title = {{The quadratic bundle of general form and the nonlinear evolution equations. II. Hierarchies of Hamiltonian structures}},
journal = {Bulg. J. Phys.}
}

@article{MikShaYam87,
year = {1987},
publisher = {},
volume = {42},
number = {4},
pages = {1-63},
author = {Mikhailov, A.~V. ~Shabat, A.~B and Yamilov, R.~I.},
title = {{The~symmetry approach to the classification of non-linear equations. Complete lists of integrable systems}},
journal = {Russian Math. Surveys}
}

@article{Mullahasanoglu:2023djn,
    author = "Mullahasanoglu, Mustafa",
    title = "{Liouville integrable binomial Hamiltonian system}",
    eprint = "2304.04500",
    archivePrefix = "arXiv",
    primaryClass = "nlin.SI",
    doi = "10.1088/1742-6596/2667/1/012041",
    journal = "J. Phys. Conf. Ser.",
    volume = "2667",
    number = "1",
    pages = "012041",
    year = "2023"
}

@article{Ashwinkumar:2023zbu,
    author = "Ashwinkumar, Meer and Sakamoto, Jun-ichi and Yamazaki, Masahito",
    title = "{Dualities and discretizations of integrable quantum field theories from 4d Chern-Simons theory}",
    eprint = "2309.14412",
    archivePrefix = "arXiv",
    primaryClass = "hep-th",
    doi = "10.4310/atmp.251118001759",
    journal = "Adv. Theor. Math. Phys.",
    volume = "29",
    number = "6",
    pages = "1509--1694",
    year = "2025"
}

@article{Costello:2019tri,
    author = "Costello, Kevin and Yamazaki, Masahito",
    title = "{Gauge Theory And Integrability, III}",
    eprint = "1908.02289",
    archivePrefix = "arXiv",
    primaryClass = "hep-th",
    reportNumber = "IPMU19-0110",
    month = "8",
    year = "2019"
}

@article{Yamazaki:2025yan,
    author = "Yamazaki, Masahito",
    title = "{Gauge Theory and Integrability: An Overview}",
    eprint = "2509.07628",
    archivePrefix = "arXiv",
    primaryClass = "hep-th",
    year = "2025"
}

@article{Polchinski:1983gv,
    author = "Polchinski, Joseph",
    title = "{Renormalization and Effective Lagrangians}",
    reportNumber = "HUTP-83-A018",
    doi = "10.1016/0550-3213(84)90287-6",
    journal = "Nucl. Phys. B",
    volume = "231",
    pages = "269--295",
    year = "1984"
}

@article{Bervillier:2004mf,
    author = "Bervillier, C.",
    title = "{The Wilson-Polchinski exact renormalization group equation}",
    eprint = "hep-th/0405025",
    archivePrefix = "arXiv",
    reportNumber = "SACLAY-SPH-T04-058, T04-058",
    doi = "10.1016/j.physleta.2004.09.037",
    journal = "Phys. Lett. A",
    volume = "332",
    pages = "93--100",
    year = "2004"
}

@article{Akhmedov:2010sw,
    author = "Akhmedov, E. T. and Musaev, E. T.",
    title = "{Exact statement for Wilsonian and holographic renormalization group}",
    eprint = "1001.4067",
    archivePrefix = "arXiv",
    primaryClass = "hep-th",
    doi = "10.1103/PhysRevD.81.085010",
    journal = "Phys. Rev. D",
    volume = "81",
    pages = "085010",
    year = "2010"
}

@article{Akhmedov:2010mz,
    author = "Akhmedov, E. T. and Gahramanov, I. B. and Musaev, E. T.",
    title = "{Hints on integrability in the Wilsonian/holographic renormalization group}",
    doi = "10.1134/S0021364011090037",
    journal = "JETP Lett.",
    volume = "93",
    pages = "545--550",
    year = "2011"
}

@article{arik:1989,
author = {Arik,Metin  and Neyzi,Fahrünisa  and Nutku,Yavuz  and Olver,Peter J.  and Verosky,John M. },
title = {{Multi‐Hamiltonian structure of the Born–Infeld equation}},
journal = {Journal of Mathematical Physics},
volume = {30},
number = {6},
pages = {1338-1344},
year = {1989},
doi = {10.1063/1.528314}
}

@article{OEIS,
     author = "N. J. A. Sloane",
     title = "{The On-Line Encyclopedia of Integer Sequences}"
}

@article{shabat,
     author = "A.B. Shabat and V.E. Adler and
V.G. Marikhin and V.V. Sokolov",
     title = "{Encyclopedia of integrable systems}"
}

@article{olver,
author = {Olver,Peter J.  and Nutku,Yavuz },
title = {Hamiltonian structures for systems of hyperbolic conservation laws},
journal = {Journal of Mathematical Physics},
volume = {29},
number = {7},
pages = {1610-1619},
year = {1988},
doi = {10.1063/1.527909}
}

@article{motzin,
    author = "R. Oste and J. Van der Jeugt",
    title = "{“Motzkin paths, Motzkin polynomials and
recurrence relations}",
    journal = "Electronic journal of combinatorics",
    volume = "22",
    number = "2",
    year = "2015"
}

@article{dubrovin,
author = { B. A.   Dubrovin  and  S. P.   Novikov },
title = {{Hamiltonian formalism of one-dimensional systems of hydrodynamic type, and the Bogolyubov-Whitman averaging method}},
pages = "382-386",
volume = "11",
journal = "World Scientific Series in 20th Century Physics",
doi = {10.1142/9789814317344_0051}
}

@book{Arnold,
    author = "V. I. Arnol’d",
    title = "Mathematical methods of classical mechanics",
    publisher = " Springer
Science \& Business Media",
    volume = "60",
    year = "2013"
}

@article{Krippendorf_2021,
	doi = {10.1002/prop.202100057},
	year = 2021,
	eprint = {2103.07475},
	archivePrefix = {arXiv},
	primaryClass = {nlin.SI},
	publisher = {Wiley},
	volume = {69},
	number = {7},
	pages = {2100057},
	author = {Sven Krippendorf and Dieter Lüst and Marc Syvaeri},
	title = {{Integrability Ex Machina}},
	journal = {Fortschritte der Physik}
}

@article{Gieres:2021ekc,
    author = "Gieres, Francois",
    title = "{Covariant canonical formulations of classical field theories}",
    eprint = "2109.07330",
    archivePrefix = "arXiv",
    primaryClass = "hep-th",
    doi = "10.21468/SciPostPhysLectNotes.77",
    journal = "SciPost Phys. Lect. Notes",
    volume = "77",
    pages = "1",
    year = "2023"
}

@article{Driezen:2021cpd,
    author = "Driezen, Sibylle",
    title = "{Modave Lectures on Classical Integrability in 2d Field Theories}",
    eprint = "2112.14628",
    archivePrefix = "arXiv",
    primaryClass = "hep-th",
    doi = "10.22323/1.404.0002",
    journal = "PoS",
    volume = "Modave2021",
    pages = "002",
    year = "2022"
}

@article{Ferapontov2025,
  author  = {Ferapontov, Evgeny V. and Vermeeren, Mats},
  title   = {Lagrangian multiforms and dispersionless integrable systems},
  journal = {Letters in Mathematical Physics},
  volume  = {115},
  number  = {6},
  pages   = {125},
  year    = {2025},
  issn    = {1573-0530},
  doi     = {10.1007/s11005-025-02016-w}
}

@article{MANGANARO2024,
title = {{Solutions to the wave equation for commuting flows of dispersionless PDEs}},
journal = {International Journal of Non-Linear Mechanics},
volume = {159},
pages = {104611},
year = {2024},
issn = {0020-7462},
doi = {https://doi.org/10.1016/j.ijnonlinmec.2023.104611},
author = {Natale Manganaro and Alessandra Rizzo and Pierandrea Vergallo}
}

@article{Kamchatnov,
    author = {Kamchatnov, A. M.},
    title = {Asymptotic integrability of nonlinear wave equations},
    journal = {Chaos: An Interdisciplinary Journal of Nonlinear Science},
    volume = {34},
    number = {11},
    pages = {113117},
    year = {2024},
    issn = {1054-1500},
    doi = {10.1063/5.0227082}
}

@article{Krynski2023,
  author = {Kry{\'n}ski, Wojciech},
  title = {{Deformations of dispersionless Lax systems}},
  journal = {Classical and Quantum Gravity},
  volume = {40},
  number = {23},
  pages = {235013},
  year = {2023},
  doi = {10.1088/1361-6382/ad0748}
}

@article{KAMCHATNOV2024,
title = {{Quasiclassical integrability condition in AKNS scheme}},
journal = {Physica D: Nonlinear Phenomena},
volume = {460},
pages = {134085},
year = {2024},
issn = {0167-2789},
doi = {https://doi.org/10.1016/j.physd.2024.134085},
author = {A.M. Kamchatnov and D.V. Shaykin}
}

@article{Ferapontov2009,
  author = {Ferapontov, E. V. and Moro, A. and Novikov, V. S.},
  title = {Integrable equations in $2+1$ dimensions: deformations of dispersionless limits},
  journal = {Journal of Physics A: Mathematical and Theoretical},
  volume = {42},
  number = {34},
  pages = {345205},
  year = {2009},
  issn = {1751-8121},
  doi = {10.1088/1751-8113/42/34/345205}
}

@article{Matthews2000,
  author = {Matthews, Keith R.},
  title = {{The Diophantine equation $x^2-Dy^2=N$, $D>0$}},
  journal = {Expositiones Mathematicae},
  volume = {18},
  number = {4},
  pages = {323--331},
  year = {2000}
}

@article{Lenstra2002,
  author = {Lenstra, Jr., Hendrik W.},
  title = {{Solving the Pell equation}},
  journal = {Notices of the American Mathematical Society},
  volume = {49},
  number = {2},
  pages = {182--192},
  year = {2002}
}

@article{Choudhuri_2007,
   title={{Lagrangian Approach to Dispersionless KdV Hierarchy}},
   ISSN={1815-0659},
   DOI={10.3842/sigma.2007.096},
   journal={Symmetry, Integrability and Geometry: Methods and Applications},
   publisher={SIGMA (Symmetry, Integrability and Geometry: Methods and Applications)},
   author={A. Choudhuri and B. Talukdar and U. Das},
   year={2007},
   }

@article{Jackiw:2004nm,
    author = "Jackiw, R. and Nair, V. P. and Pi, S. Y. and Polychronakos, A. P.",
    title = "{Perfect fluid theory and its extensions}",
    eprint = "hep-ph/0407101",
    archivePrefix = "arXiv",
    reportNumber = "MIT-CTP-3509 BUHEP-04-07",
    doi = "10.1088/0305-4470/37/42/R01",
    journal = "J. Phys. A",
    volume = "37",
    pages = "R327--R432",
    year = "2004"
}

@article{KunduStramppOevel1995,
  author  = {Kundu, Anjan and Strampp, Walter and Oevel, Walter},
  title   = {Gauge transformations of constrained {KP} flows:
             New integrable hierarchies},
  journal = {Journal of Mathematical Physics},
  volume  = {36},
  number  = {6},
  pages   = {2972--2984},
  year    = {1995},
  doi     = {10.1063/1.531336}
}

@book{Faddeev:1987ph,
    author = "Faddeev, L. D. and Takhtajan, L. A.",
    title = "{Hamltonian methods in the theory of solitons}",
    year = "2007",
    doi= "10.1007/978-3-540-69969-9",
    publisher= "Springer Berlin, Heidelberg"
}

@article{Dunajski:2008at,
    author = "Dunajski, Maciej",
    editor = "Barlow, Roger",
    title = "{Interpolating Dispersionless Integrable System}",
    eprint = "0804.1234",
    archivePrefix = "arXiv",
    primaryClass = "nlin.SI",
    reportNumber = "PREPRINT-DAMTP-2008-27",
    doi = "10.1088/1751-8113/41/31/315202",
    journal = "J. Phys. A",
    volume = "41",
    pages = "315202",
    year = "2008"
}

@article{Bonkile:2018bjm,
    author = "Bonkile, Mayur P. and Awasthi, Ashish and Lakshmi, C. and Mukundan, Vijitha and Aswin, V. S.",
    title = "{A systematic literature review of Burgers{\textquoteright} equation with recent advances}",
    doi = "10.1007/s12043-018-1559-4",
    journal = "Pramana",
    volume = "90",
    number = "6",
    pages = "69",
    year = "2018"
}

@article{Sipka_2023,
doi = {10.1088/1751-8121/ad0803},
year = {2023},
publisher = {IOP Publishing},
volume = {56},
number = {49},
pages = {495201},
author = {Šípka, Martin and Pavelka, Michal and Esen, Oğul and Grmela, Miroslav},
title = {Direct Poisson neural networks: learning non-symplectic mechanical systems},
journal = {Journal of Physics A: Mathematical and Theoretical}
}

@article{Reshetikhin:2015eja,
    author = "Reshetikhin, Nicolai and Sridhar, Ananth",
    title = "{Integrability of Limit Shapes of the Six Vertex Model}",
    eprint = "1510.01053",
    archivePrefix = "arXiv",
    primaryClass = "math-ph",
    doi = "10.1007/s00220-017-2983-x",
    journal = "Commun. Math. Phys.",
    volume = "356",
    number = "2",
    pages = "535--565",
    year = "2017"
}

@article{Lou:2024fhu,
    author = "Lou, Senyue",
    title = "{Progresses on Some Open Problems Related to Infinitely Many Symmetries}",
    eprint = "2406.00208",
    archivePrefix = "arXiv",
    primaryClass = "nlin.SI",
    doi = "10.3390/math12203224",
    journal = "Mathematics",
    volume = "12",
    number = "20",
    pages = "3224",
    year = "2024"
}

@article{Calderbank:2016bqu,
    author = "Calderbank, David M. J. and Kruglikov, Boris",
    title = "{Integrability via geometry: dispersionless differential equations in three and four dimensions}",
    eprint = "1612.02753",
    archivePrefix = "arXiv",
    primaryClass = "math.AP",
    doi = "10.1007/s00220-020-03913-y",
    journal = "Commun. Math. Phys.",
    volume = "382",
    number = "3",
    pages = "1811--1841",
    year = "2021"
}

@article{Ismailov2011,
  author  = {Ismailov, Mansur I.},
  title   = {Integration of nonlinear system of four waves with two
             velocities in $(2+1)$ dimensions by the inverse scattering
             transform method},
  journal = {Journal of Mathematical Physics},
  volume  = {52},
  number  = {3},
  pages   = {033504},
  year    = {2011},
  doi     = {10.1063/1.3560476}
}

@article{Fan2001,
  author  = {Fan, Engui},
  title   = {{A family of completely integrable multi-Hamiltonian
             systems explicitly related to some celebrated equations}},
  journal = {Journal of Mathematical Physics},
  volume  = {42},
  number  = {9},
  pages   = {4327--4344},
  year    = {2001},
  doi     = {10.1063/1.1389288}
}

\end{document}